# Time dependent singularities in incompressible potential flow

Daniel Weihs

*Department of Aerospace Engineering and Autonomous Systems Program*

*Technion, Haifa 32000, Israel*

Abstract

The flow-field resulting from a starting source –type singularity is presented. This is a generalization of the classical source in potential flow to time dependent situations, including sources in oncoming flow. This model is applicable to initial spreading of oil on the surface of the sea resulting from an underwater leak

Introduction

The use of singularities is one of the fundamental building blocks of potential flow theory. Singularities such as sources, sinks, doublets etc. are used to describe flows and body shapes by applying the no-penetration condition and to calculate forces and moments on those bodies[1]. However, these all deal with steady (time-independent) flow. In this note, we look at the flows (and transient fluid body shapes) resulting from these singularities varying in time, and specifically, starting up at a given time in an incompressible fluid. As a result, the fluid mass emitted grows with time and causes the formation of a growing disturbance. In order to compare with existing solutions of time independent flow, we assume that the fluid emitted from the source and the ambient fluid, do not mix, to show how the Rankine bodies formed by steady flow are approached. There are intrinsic problems with studying time dependent flow in incompressible fluid, as the speed of propagation is infinite, and streamlines do not exist- which may explain why this problem has not been discussed previously. We circumvent these issues by following pathlines and calculation of volume emitted. This type of analysis is the basis for calculating the spread of oil from a surface leak in time, however several additional factors such as surface tension, wind –driven waves and mixing, etc. need to be added.

## I. Analysis

We start by looking at the simplest singularity, a two dimensional point source of incompressible fluid in a vacuum.

From the classical definition for a steady incompressible and inviscid two-dimensional source located at the origin we have a potential[1],

$$\phi = -Q \ln r \quad (1)$$

In a cylindrical coordinate system, where Q is the mass flow rate. In the following unit density is assumed. The local velocity is purely radial

$$v_r = v = kQ/r \quad (2)$$

This is obviously a mathematical idealization. The physical realization of such a source would be a small, but finite circular orifice of radius $R_0$ and large width ( normal to the plane of fluid emission) . At radius $R_0$, due to the uniform circumferential distribution of mass efflux there will be an emerging radial velocity of

$$V_{R0} = Q/2\pi R_0 \quad (3)$$

Now let's look at the same source, but with the flow starting at t=0. For this first example we take a step function of time, i.e. $V_{R0}=0$ for t<0 and $v=V_{R0}$ @ $r=R_0$ and $0 \leq t < \infty$

Assuming now that the exiting fluid has unit density and moves at zero gravity, we want to determine the flow-field formed. This, trivially, has a circular shape growing in time with outer border $R_t$ defined by the total mass emitted up to that time.

$$R_t = \sqrt{\frac{Qt}{\pi} + R_0^2} \quad (4)$$

A similar process leads to growing spherical shapes for the three dimensional source. We can look at the three-dimensional case too, as we do not invoke streamlines in the time dependent flow.

Next, and more interestingly, take the case of this starting source in a uniform flow of speed U of fluid of the same density. The steady potential now is[1]

$$\phi = -Q \ln r + Ur \sin \theta \quad (5)$$

For which one obtains a two dimensional Rankine shape in steady flow. Translated into our problem, the Rankine shape will be obtained at large (infinite) times. As mentioned above, the distance from the origin (source position) to the stagnation point in time-independent flow is $L_s = Q/U$ in the two dimensional case and ( or, as can be shown ), is $\sqrt{Q/U}$ for the axisymmetric case , respectively. This allows us to define a characteristic length to be used in normalizing coordinates.

Unfortunately, we cannot directly use Rankine's method, which involves streamlines as we now have time dependent flow in which streamlines do not exist as such. In this case, there are geometrical parameters which are obtained from the mass flux Q and the oncoming speed U. These are the width of the body at infinity, and the distance between the stagnation point produced at an upstream distance from the source, under steady conditions. We then choose an arbitrary $R_0$ which should be much smaller than the Rankin body asymptotic width $L_s$.

As the fluid is incompressible and inviscid, if we do not allow mixing we can define the size of the emerging area by integrating the mass efflux over time. The present example requires the mass flux rate to be constant with time, starting instantaneously at t=0. The stagnation point upstream of the source is formed therefore at the time–independent location instantaneously, while the rest of the flow, which would have had circular form, is now deformed in the downstream direction by the oncoming flow. In order to calculate the shape we look at the time trajectory of particles belonging to the flow emitted at $\theta=0^+$, i.e. very close to the upstream direction.

The second case under examination is the three dimensional axisymmetric case. This case is similar to the 2D case if the definition of $R$ is as follows

$$R = \sqrt{y^2 + z^2} \quad (6)$$

A three dimensional source in a uniform stream gives the velocities

$$u_r = \frac{dR}{dt} = U\cos\theta + \frac{Q}{R^2} \quad (7)$$

$$u_\theta = \frac{Rd\theta}{dt} = -U\sin\theta \quad (8)$$

Then in this case the path line equation is

$$\frac{UR^2 \sin^2\theta}{2} - Q\cos\theta = C \quad (9)$$

where $C$ is an integration constant. Finding the position of the particles can be done by integrating eq. (8) to get

$$-\int_{\theta_0}^{\theta_1} \frac{R}{U\sin\theta}d\theta = \int_{\theta_0}^{\theta_1} \frac{\sqrt{2(C+Q\cos\theta)}}{U^{3/2}\sin^2\theta}d\theta = t \quad (10)$$

for a given time $t$ a value of $\theta_1$ is sought. This is achieved by a bi-section procedure with respect to $\theta$ for the integration in eq. (10). After determining $\theta_1$, substitution in eq. (9) gives $R_1$. Then, running the procedure for $0 \leq \theta_0 \leq \pi$ gives the requested shape.

To illustrate the time development of the source efflux we determine the values of Q in the case of "step" source intensity in terms of the oncoming flow. We choose the ratio of Q/U (in the two dimensional case) and $\sqrt{Q/U}$ (in the three dimensional case) to be $O(10^{-1})$. While the two-dimensional model assumes infinite depth, this analysis can describe finite (even narrow) two dimensional layers (still within the potential flow regime) by defining the flow per unit depth of the efflux.

Results

In order to calculate the form of the body of fluid produced by the time-dependent source, we assume the source to have a finite radius . To establish this finite radius so that the flow is close enough to that of a point source, we uset the distance Q/U of the asymptotic solution ( (the Rankine body). This length is a yardstick for the minimum radius we use to calculate the efflux, to avoid the singularity at the origin. The minimum radius is thus chosen as $R_0 << Q/U$
The results of this calculation appear in Figures 1 ( the two-dimensional case) and
2 (axisymmetric case).

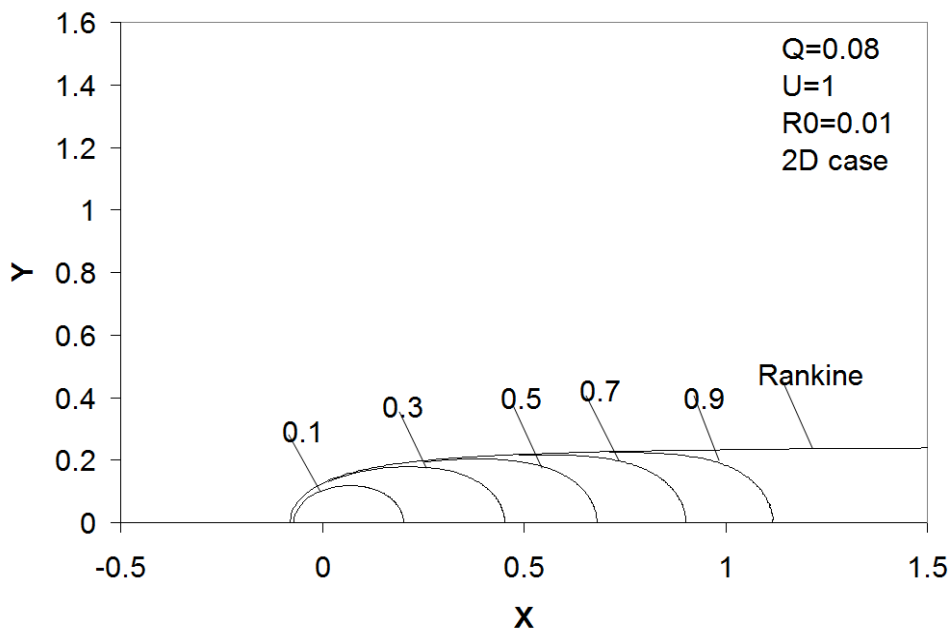

**Figure 1. The shape produced by a source in oncoming flow in two dimensions at different nondimensional times. The numbers relate to the nondimensional time elapsed since emission started**.

The values for the source strength Q , and the nondimensional times $T=tU^2/\pi Q$ were chosen arbitrarily, to clearly showcase the development of the body of fluid emitted by the source. A similar calculation is made for the axisymmetric three dimensional flow, where the body produced asymptotically approaches the so-called Rankine solid (see Figure 2).

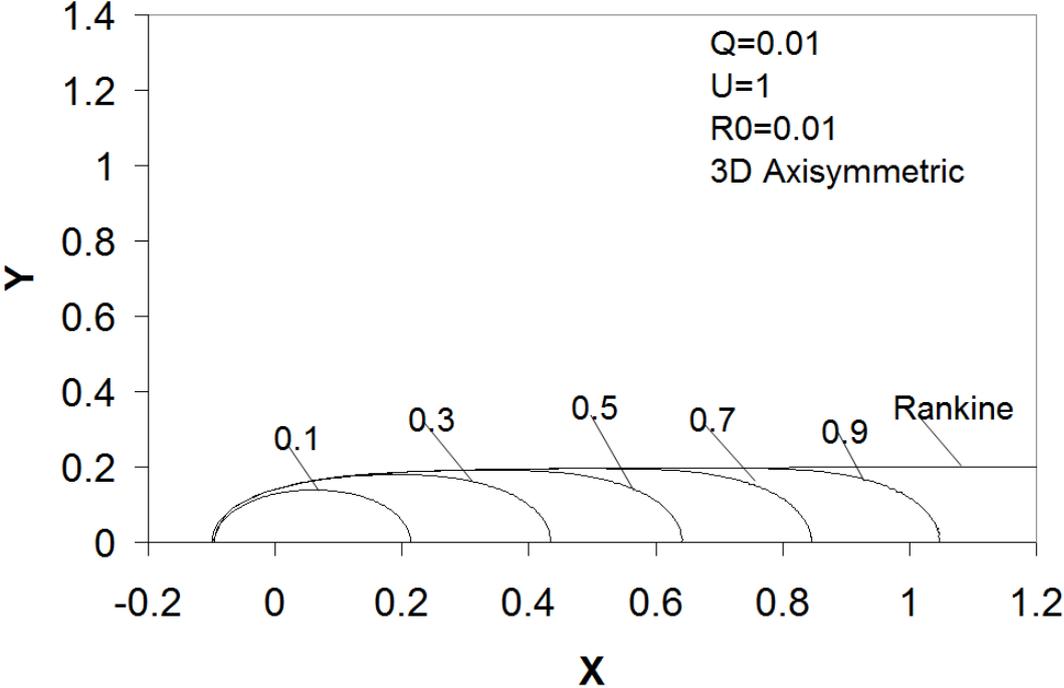

**Figure 2. The shape produced by a source in oncoming flow in Axisymmetric three dimensional flow**.

Next we look at a time dependent growth of the efflux, assuming a linear growth of source strength with time, up to the steady value.

$$Q(t)=Mt \quad \text{for} \quad 0<t<1 \text{ and } Q(t) =M \text{ for } t>1$$

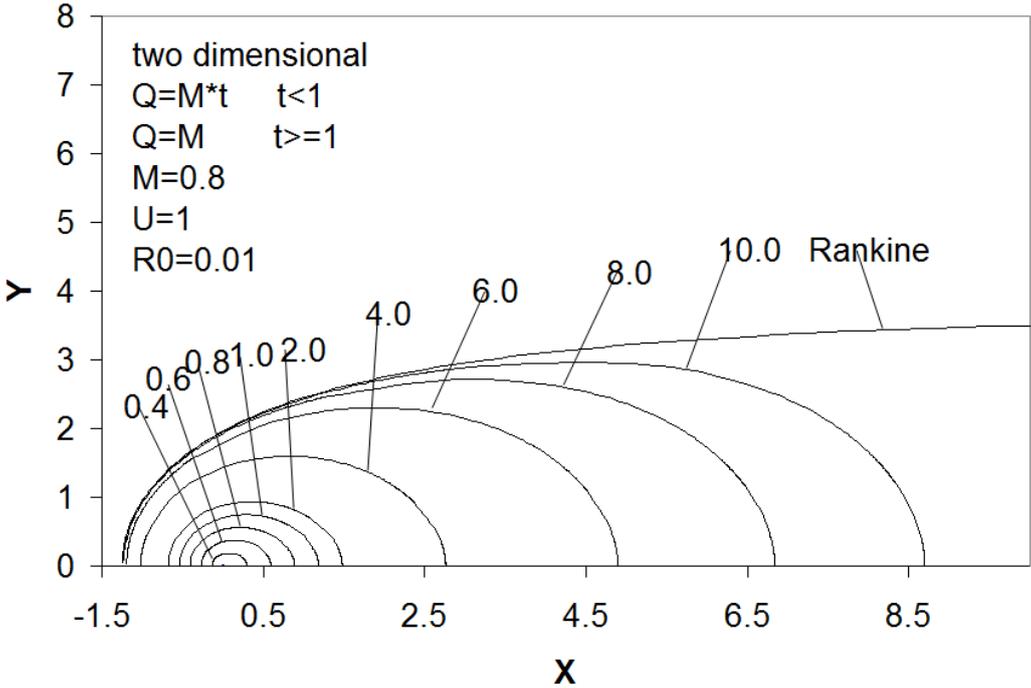

**Fig 3 . Efflux shape of linearly strengthening , two dimensional source in oncoming flow**.

The same calculation is now performed for the three-dimensional source

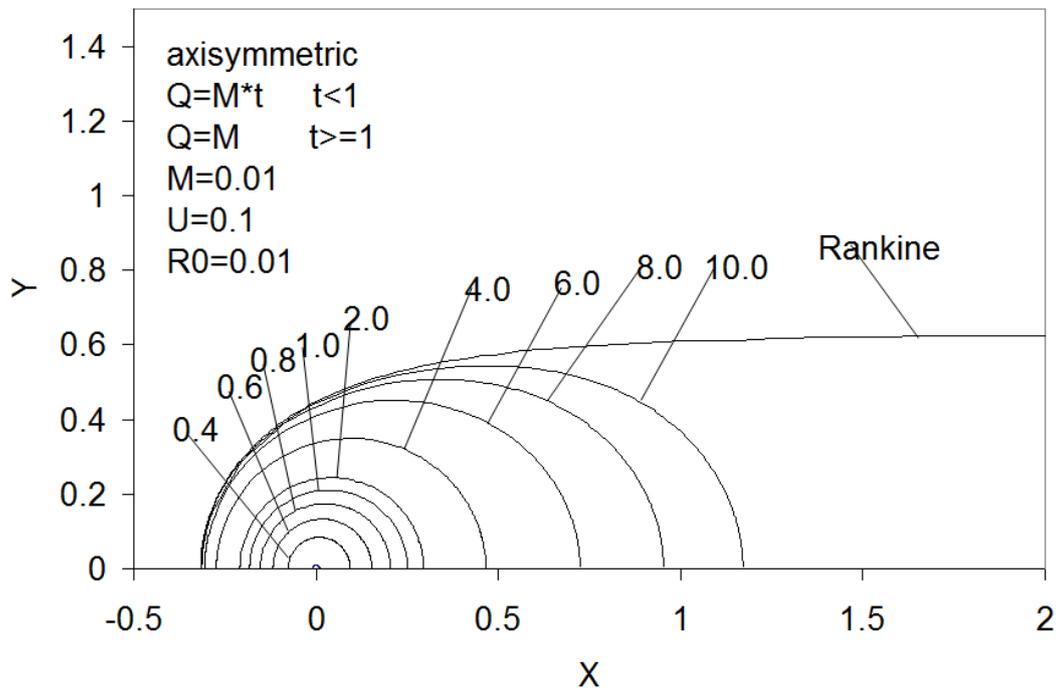

**Fig.5.  Axisymmetric Linearly growing Source in Oncoming flow**

Discussion

We obtain the shape of the time dependent efflux of sources in two and three dimensional flow-fields, modelling the shape of an oil slick initially spreading on the ocean surface , including the effects of a tide or other driven currents. Axisymmetric solutions are also presented, for spread of point sources in uniform flow. The shapes obtained do not have any inflection points, so no instability is expected for infinitesimal perturbations. Further possibilities have been caculated, such as time-varying oncoming flows but no unexpected features appeared.

Finally we look at the time development of a body, such as an oil slick, that was produced in finite time and stopped, following it's trajectory downstream due the prevailing currents.

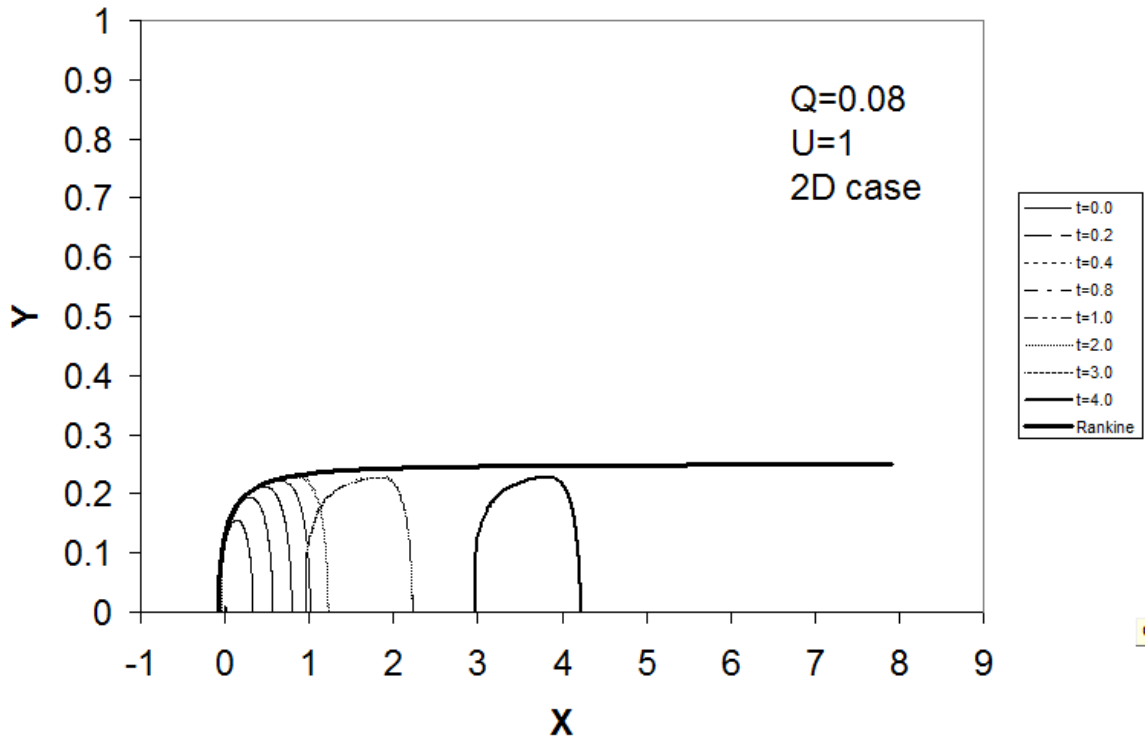

We see here that the body of fluid essentially keeps it's shape, slowly lengthening and narrowing . The figure is laterally distorted , to show some development in time. At the times mentioned, one expects viscous forces to become important and cause instability of the interface. Such effects are out of the range of the present paper, and we are studying these viscous changes at present.

## Statements

Conflict of Interest: The author declares that he has no conflict of interest.
Funding: There is no funding source.
The manuscript does not contain clinical studies or patient data.

## Acknowledgements

The author wishes to thank Dr. Gal Davidi for performing the calculations shown in the figures.